**12**

# FIRST PHOTOMETRIC ANALYSES AND CLASSIFICATION FOR THE W UMA TYPE BINARY SYSTEMS GSC 2362-2866 AND GSC 107-596

A. Essam[*], N. S. Awadalla[*],
M. A. El-Sadek[**], and G. B. Ali[*]

**ABSTRACT:** *The CCD observations for the W UMa type binary systems GSC 2362-2866 and GSC 107-596 have been analyzed using the Wilson-Devinney Code to determine their photometric orbital and physical parameters. The results show that both systems may be classified as A-subtype of W-UMa eclipsing binary. The mass ratio of GSC 2362-2866 is found to be q = 0.73, with an over-contact degree of f = 0.122%. While the mass ratio of the system GSC 107-596 is found to be q = 0.70 with an over-contact degree of f = 0.166% in V_band and q = 0.69 and f = 0.111% in R band*

## INTRODUCTION

The contact binary system GSC 2362-2866 (USNO–A2.0 1200-01935713 = NSVS 6713581), which have the coordinates $\alpha(2000) = 04^h 03^m 41^s$, $\delta(2000) = +32° 27' 6"$, was discovered to be variable by Tommi Itkonen and Pertti Pääkkönen on December 2005, using CCD camera, attached on Cassegrain, of Jakokoski observatory, Finland (see Olah 2006). The system was classified as W UMa eclipsing binary, with a period $0.29768^d = 7.144^h$, and the magnitude ranging from $13.45^m$ - $14.05^m$ in V band with epoch of Min I equals 2453756.34206.

The other contact binary system GSC 107-596 (ASAS J050837+0512.3 =NSVS 12310076 = UCAC2 33518439), which have the coordinates $\alpha(2000) = 05^h 08^m 36.45^s$, $\delta(2000) = +5° 12' 23.1"$, was discovered to be variable by Blättler, E. and Diethelm, R. on March 2007 (see Olah 2007), using 0.15-m Starfire refractor (Private observatory) in Wald, Switzerland. The system having $B_{mag}=11.9$ and $R_{mag}=10.4$, it was classified as W UMa eclipsing binary, with a period $0.2663496d = 6.392^h$, and the epoch of Min I equals 2454066.4302.

## OBSERVATIONS

___________________________________________________________
*\*National Research Institute of Astronomy and Geophysics, Helwan, Cairo, Egypt*
*\*\*Astronomy and Space Science Dept., Kuwait Science Club, Kuwait*

One set of non-analyzed V-band observational data for the contact binary system GSC 2362-2866 has been observed by Pertti Paakkonen and Tommi Itkonen during the time interval between 22-12-2005 and 20-10-2006. The CCD camera, SBIG ST-1001E, attached on Cassegrain, of Jakokoski telescope; and The CCD camera, SBIG STL-1001E, attached to RCT of Hankasalmi telescope, Finland, have been used to make these observations.

The phase curve is based on 395 observations. They used the stars GSC 2362-2188, V=10.928; GSC 2362-2442, V=11.422; GSC 2362-2773, V=10.934 as comparison stars and the star GSC 1804-2485 as check star.

The mean epoch of minimum light was determined from the primary and secondary eclipses using the results of parabolic fits. The Heliocentric Julian Date of the primary minima can be computed by the following formula (see Olah 2006).

HJD Min I = 2453756.34206 + 0.29768d * E

Where E is the number of integer cycles. This ephemeris is used to calculate the phases and derives the normal points of the light curve ΔV (Differential Magnitude), for all observations, see Figure 1.

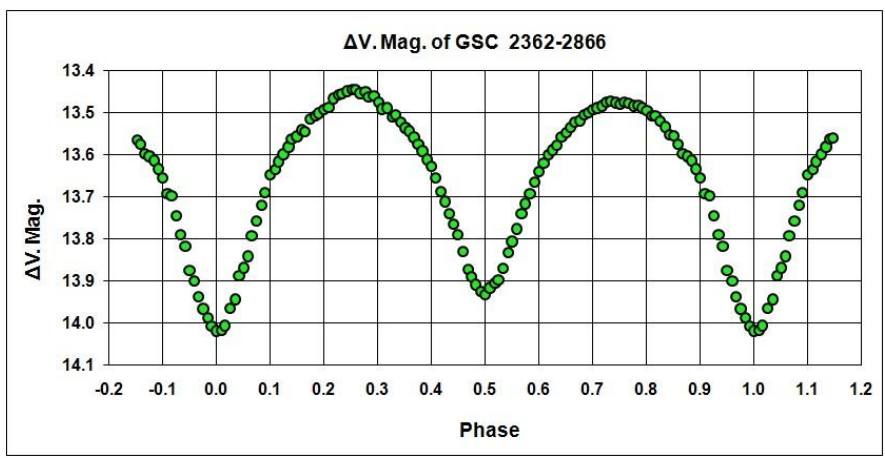

Fig. 1: Normal points light curve of the V-band for GSC 2362-2866

The other sets of non-analyzed CCD observational data in V and R bands for the contact binary system GSC 107-596 have been observed by

Blättler, E. using SBIG ST-7 camera attached to 0.15-m Starfire refractor in Wald, Switzerland. The observations were made during 5 nights between JD 2454066 and JD 2454114. A total of 221 measurements in both colours were obtained, using GSC 107-1120 as comparison and GSC107-165 as check star. A linear regression of the 16 times of minima with the ROTSE1 data yields the following ephemeris (see Olah 2007).

$$\text{HJD Min I} = 2454066.4302 + 0.2663496d * E$$

where E is the number of integer cycles. This ephemeris is used to calculate the phases and derives the normal points of the light curves ΔV and ΔR (Differential Magnitude) for all observations, see Figures 2 and 3.

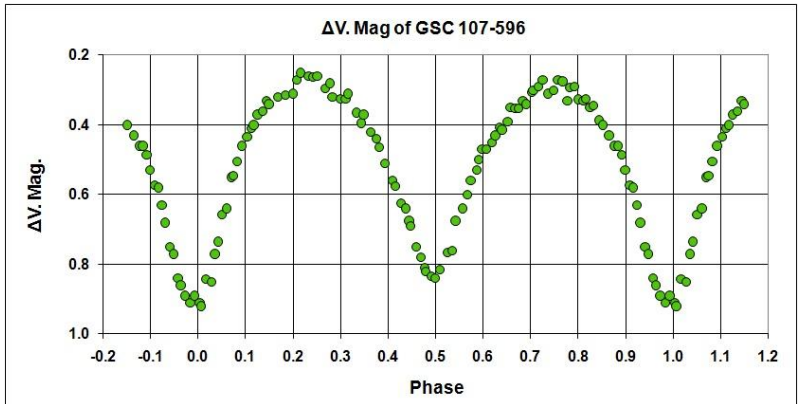

**Fig. 2: Normal points light curve of the V-band for GSC 107-596**

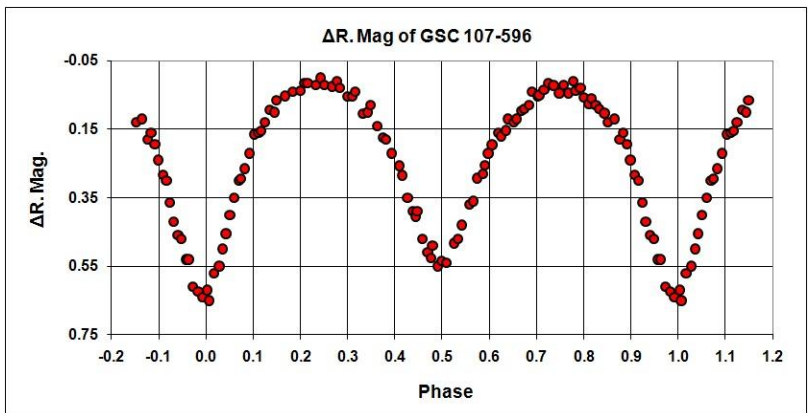

**Fig. 3: Normal points light curve of the R-band for GSC 107-596**

**THE PHOTOMETRIC DATA ANALYSIS OF GSC 2362-2866**

In order to determine the parameters of the system, we used the Wilson and Devinney (WD hereafter) Differential Correction program (Wilson et al., 2003). The model has been described and quantified in papers by Wilson & Devinney (1971), Wilson (1979, 1990, 1993), and Wilson &Van Hamme (2003) that include its main theory, organization, and concepts, as well as much of the mathematics.

The analyses were done for the available V light curve (see Figure 1) after transfer the magnitude to normalized flux (N. Flux). The light curve appears to exhibit a typical O'Connell effect (O'Connell 1951), known in many eclipsing binaries and explained by spot activity on the components (Zhai et al, 1988).

Different solutions, with and without spot/s on the components are tested. The preliminary solutions were determined by means of the synthetic light-curve program. After many trials we derived a set of parameters which marginally represent the observed light curve. The exponent of gravity darkening, $g_1 = g_2 = 0.32$ (Lucy 1967) and bolometric albedo, $A_1 = 0.5$ (Rucinski 1969) for late spectral type stars are assumed. The values of bolometric limb darkening, $x_1$ and $x_2$, are taken from Van Hamme Tables (1993). A cool spot on star 1 is supposed to treat the asymmetry of the light curve. The overcontact mode 3 of WD code is also used.

Some parameters are assumed, like, surface temperature of star 1 ($T_1$), orbital inclination (i), surface potential of two components, ($\Omega_1=\Omega_2$), bolometric albedo ($A_1$), the monochromatic luminosity of star 1 ($L_1$) and some spot parameters. The other parameters, like, phase shift, surface temperature of star 2 ($T_2$), bolometric albedo ($A_2$), monochromatic limb darkening of star 2 ($x_2$), and mass ratio (q) are adjusted and employed in the second part of WD program (Differential Correction program; DC). The relative brightness of the star 2 was calculated by the stellar atmosphere model.

The parameters derived are listed in Table (1) with their standard deviation. The final fit of the observations was plotted in Figure (4) for V filter. The star1 is the more massive and hotter component, while star 2 is the less massive and cooler one.

Using the geometrical and physical parameters listed in Table (1) with the Binary Maker 3 program (Bradstreet 2005), to produce the Roche geometry of the system in Figure (5), which shows the degree of contact. The same parameters are employed with the same program to display the system configuration at different phases (0.0, 0.25, 0.5, and 0.75) with the positions of the spots as shown in Figure (6). The resultant fits seem to be quite satisfactory.

**Table 1: The orbital and physical parameters of the system GSC 2362-2866**

| Parameters | Element's values for the V_band observations of 2005 & 2006 |
|---|---|
| Wavelength = $\lambda$ | 5500 Å |
| Phase Shift | $0.0008 \pm 0.0003$ |
| Inclination (i) | 73°.150 (assumed) |
| Surface Temp. $T_1$ | 5280 °K (assumed) |
| Surface Temp. $T_2$ | 5068 °K $\pm$ 3.4 °K |
| Surface potential ($\Omega_1 = \Omega_2$) | 3.242424 (assumed) |
| Mass Ratio (q) | $0.7284 \pm 0.0017$ |
| fillout parameter ($f_1$ & $f_2$) | 0.12212 % |
| Reflection= ($A_1$) | 0.500 (fixed) |
| Reflection= ($A_2$) | $0.811 \pm 0.047$ |
| Gravity exponents ($\alpha_1 = \alpha_2$) | 0.320 (fixed) |
| Angular Rotation ($F_1 = F_2$) | 1.000 |
| Volume 1 | 0.300 |
| Volume 2 | 0.195 |
| Limb Darkening $x_1$ | 0.675 (fixed) |
| Limb Darkening $x_2$ | $0.714 \pm 0.008$ |
| $L_1 / (L_1 + L_2)$ | 0.63363 (assumed) |
| $L_2 / (L_1 + L_2) = 1 - L_1$ | 0.36637 |
| $r_1$(back) | $0.449163 \pm 0.000550$ |
| $r_2$(back) | $0.396232 \pm 0.000769$ |
| $r_1$(side) | $0.414721 \pm 0.000318$ |
| $r_2$(side) | $0.358532 \pm 0.000423$ |
| $r_1$(pole) | $0.391362 \pm 0.000231$ |
| $r_2$(pole) | $0.340892 \pm 0.000325$ |
| Mean Radius 1 | 0.416339 |
| Mean Radius 2 | 0.360752 |
| Surface Area 1 | $2.205383 \pm 0.0000$ |
| Surface Area 2 | $1.659056 \pm 0.0000$ |
| **Spot of star 1** | |
| Co-Latitude | 1.46608 (assumed) |
| Longitude | $1.61989 \pm 0.02723$ |
| Spot Radius | 0.20944 (assumed) |
| Temp Factor | 0.78000 (assumed) |
| L.C. Residual | 0.005012 |

By comparing the properties of the system GSC 2362-2866 in Table (1) with the properties of A and W subtypes of the W UMa binaries (see

Rucinski 1973) through its mass ratio, degree of contact, radii, temperature difference between two components, and the transit primary minimum, we found that the system is likely to be an A-subtype of W UMa binary stars.

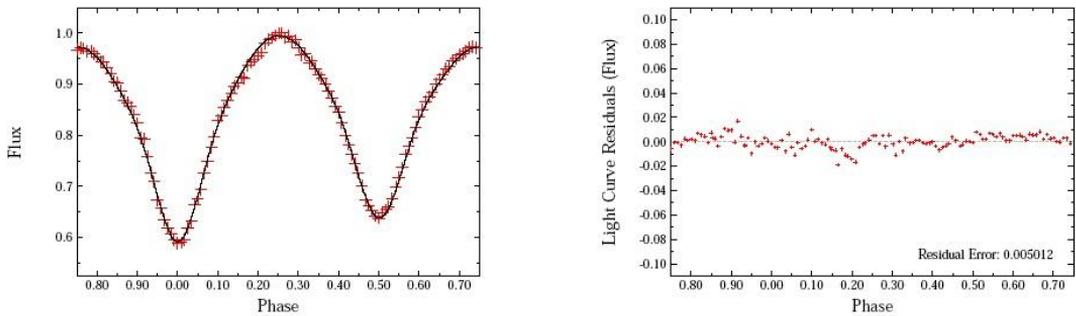

**Fig. 4: V light Curve of GSC 2362-2866 (crosses) together with their fitting (solid line) in left panel, while the light curve residual shown in right panel**

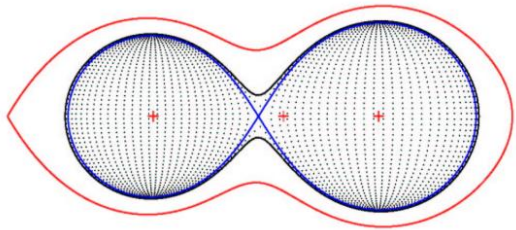

**Fig. 5: Roche Geometry of the system GSC 2362-2866 in the V-band $\Omega_1 = \Omega_2 = 3.242424$, Overcontact = 0.122 %**

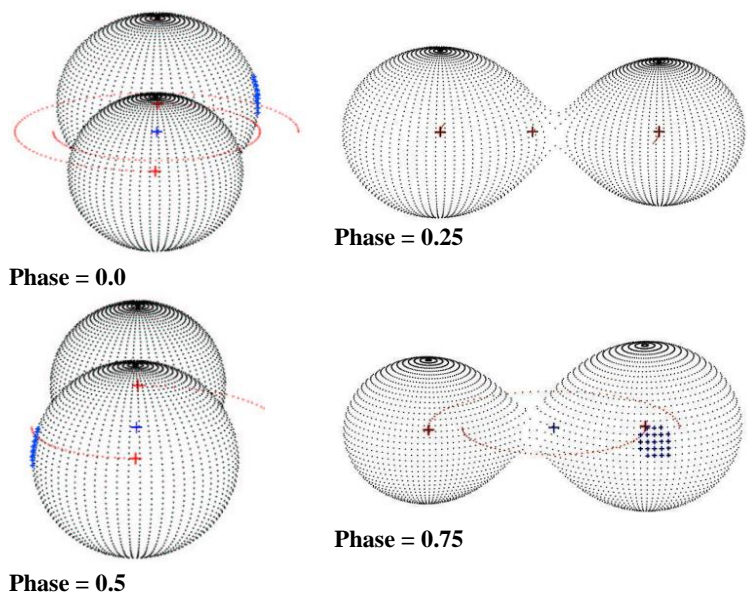

**Fig. 6: The shape of the system GSC 2362-2866 at phases 0.0, 0.25, 0.50, and 0.75**

## THE PHOTOMETRIC DATA ANALYSIS OF GSC 107-596

In order to determine the parameters of this system we used the second part of WD program (Differential Correction program), see Wilson 2003. The model has been described and quantified in papers by Wilson & Devinney (1971), Wilson (1979, 1990, 1993), and Wilson &Van Hamme (2003) that include its main theory, organization, and concepts, as well as much of the mathematics.

The analyses were done for the available V and R light curves in Figures 2 and 3, after transfer the magnitude to normalized flux (N. Flux). The light curve of this system appears to exhibit a typical O'Connell effect (O'Connell 1951). We tested different solutions, with and without spot/s on the stars. After many trials the parameters of this system are derived and found to be matched with the observed light curve.

The exponent of gravity darkening, $g_1 = g_2 = 0.32$ (Lucy 1967) and bolometric albedo, $A_1 = A_2 = 0.5$ (Rucinski 1969). The values of the bolometric limb darkening, $x_1$ and $x_2$, are taken from Van Hamme Tables (1993). A cool spot on star 1 has been supposed to treat the asymmetry of the light curve. The overcontact mode 3 of WD code is also used.

Some parameters are assumed, like, inclination (i), surface temperature of star 1 ($T_1$), orbital inclination (i), surface potential of the two components, ($\Omega_1=\Omega_2$), the monochromatic luminosity of star 1 ($L_1$) and some spot parameters.

The other parameters, like, phase shift, surface temperature of star 2 ($T_2$), monochromatic limb darkening of two components ($x_2$, $x_2$), mass ratio (q), and spot longitude are adjusted and employed in the second part of WD program (Differential Correction program; DC). The relative brightness of the star 2 was calculated by the stellar atmosphere model.

In Table 2, the derived parameters with their standard deviation are listed for the V and R bands. The final fit is plotted in Figures 7 and 10 for the V and R band respectively. The star1 indicates the more massive and hotter component, while star 2 is the less massive and cooler one.

Using the geometrical and physical parameters of V and R bands presented in Table (2) with the Binary Maker 3 program (Bradstreet 2005) the Roche geometry of the system GSC 107-596 (Figures 8 and 11) which show the degree of contact in both V and R bands respectively were produced. The same parameters in Table (2) for V and R band have been employed with the same program to display the system configuration at different phases (0.0, 0.25, 0.5, and 0.75) with the positions of the spots (see Figures 9 and 12). The resultant fitting for V and R bands seem to be quite satisfactory.

By comparing the properties of the system GSC 107-596 in Table (2) with the properties of A and W subtypes of the W UMa binaries (see Rucinski 1973) through its mass ratio, degree of contact, radii, temperature difference between two components, and the transit primary minimum, we found that the system is likely to be an A-subtype of W UMa binary stars.

**Table 2: the orbital and physical parameters of the system GSC 107-596**

| Parameters | Element's values for the V_band observations of March 2007 | Element's values for the R_band observations of March 2007 |
|---|---|---|
| Wavelength = $\lambda$ | 5500 Å | 7000 Å |
| Phase Shift | $-0.0099 \pm 0.0005$ | $-0.0072 \pm 0.0006$ |
| Inclination (i) | 75°.400 (assumed) | 75°.400 (assumed) |
| Surface Temp. $T_1$ | 3780° K (assumed) | 3810 °K ± 1.3 °K |
| Surface Temp. $T_2$ | 3669° K ± 3.7 °K | 3663 °K ± 1.3 °K |
| Surface potential ($\Omega_1 = \Omega_2$) | 3.186269 (assumed) | 3.186269 (assumed) |
| Mass Ratio (q) | $0.7056 \pm 0.0041$ | $0.6926 \pm 0.0067$ |
| fillout parameter ($f_1$ & $f_2$) | 0.16555 % | 0.110592 % |
| Reflection = ($A_1$) = ($A_2$) | 0.500 (fixed) | 0.500 (fixed) |
| Gravity exponents ($\alpha_1 = \alpha_2$) | 0.320 (fixed) | 0.320 (fixed) |
| Angular Rotation ($F_1 = F_2$) | 1.000 | 1.000 |
| Volume 1 | 0.312941 | 0.307256 |
| Volume 2 | 0.195049 | 0.186165 |
| Limb Darkening $x_1$ | $0.764 \pm 0.024$ | $0.767 \pm 0.061$ |
| Limb Darkening $x_2$ | $0.710 \pm 0.019$ | $0.624 \pm 0.061$ |
| $L_1 / (L_1 + L_2)$ | 0.65749 (assumed) | $0.6395 \pm 0.0059$ |
| $L_2 / (L_1 + L_2) = 1 - L_1$ | 0.34251 (calculated) | 0.3605 (calculated) |
| $r_1$(back) | $0.454881 \pm 0.001394$ | $0.454881 \pm 0.002273$ |
| $r_2$(back) | $0.395428 \pm 0.002044$ | $0.395428 \pm 0.003333$ |
| $r_1$(side) | $0.420022 \pm 0.000802$ | $0.420022 \pm 0.001309$ |
| $r_2$(side) | $0.356673 \pm 0.001107$ | $0.356673 \pm 0.001805$ |
| $r_1$(pole) | $0.395865 \pm 0.000580$ | $0.395865 \pm 0.000945$ |
| $r_2$(pole) | $0.339029 \pm 0.000850$ | $0.339029 \pm 0.001386$ |
| Mean Radius 1 | 0.422611 | 0.419748 |
| Mean Radius 2 | 0.361597 | 0.355330 |
| Surface Area 1 | 2.273965 | 2.232355 |
| Surface Area 2 | 1.654645 | 1.603022 |
| **Spot of star 1** | | |
| Co-Latitude | 1.04720 (assumed) | 1.04720 (assumed) |
| Longitude | $1.99977 \pm 0.05040$ | $2.08467 \pm 0.07033$ |
| Spot Radius | 0.20944 (assumed) | 0.20944 (assumed) |
| Temp Factor | 0.80000 (assumed) | 0.80000 (assumed) |
| L.C. Residual | 0.058535 | 0.057435 |

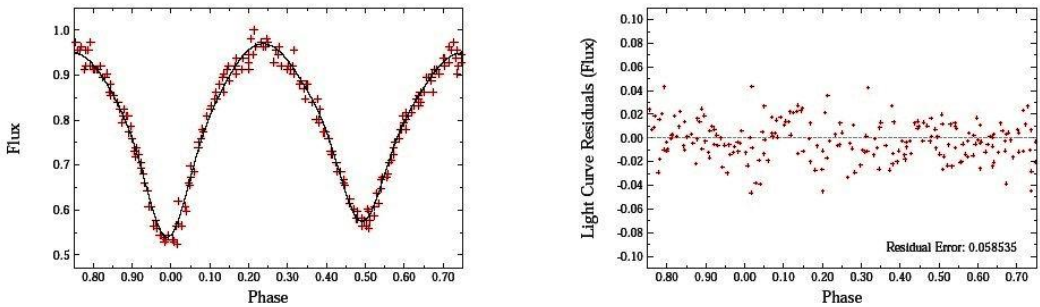

**Fig. 7: V light Curve of GSC 107-596 (crosses) together with their fitting (solid line) in left panel, while the light curve residual shown in right panel**

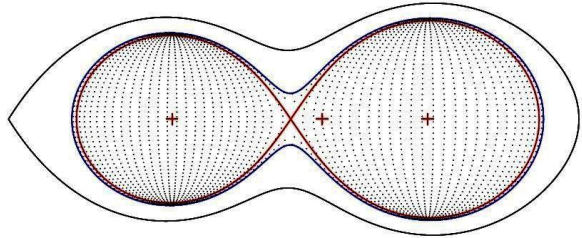

**Fig. 8: Roche Geometry of the system GSC 107-596 in the V-band $\Omega_1 = \Omega_2 = 3.186269$, Overcontact = 0.166 %**

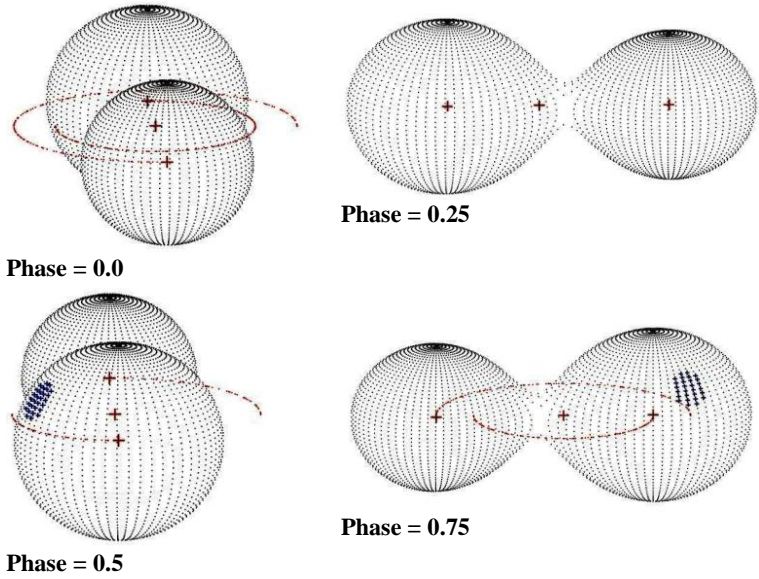

**Fig. 9: The shape of the system GSC 107-596 at phases 0.0, 0.25, 0.50, and 0.75**

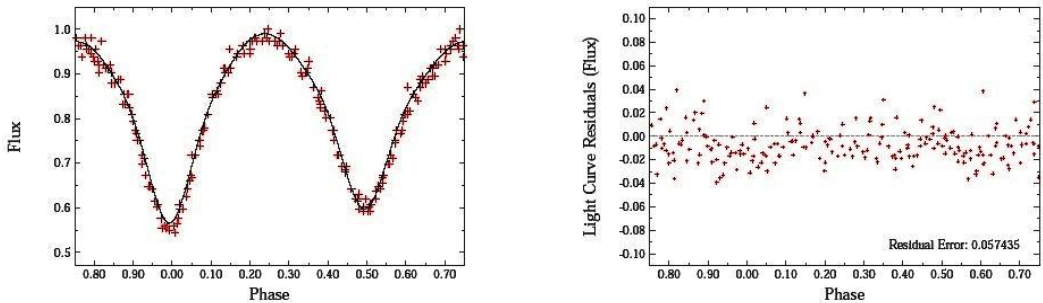

**Fig. 10: R light Curve of GSC 107-596 (crosses) together with their fitting (solid line) in left panel, while the light curve residual shown in right panel**

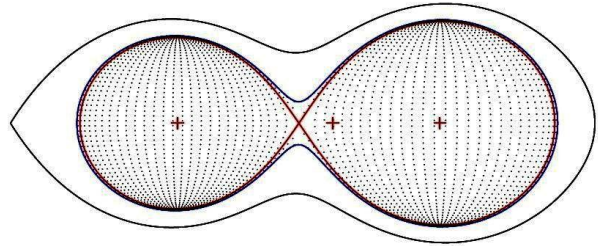

**Fig. 11: Roche Geometry of the system GSC 107-596 in the R-band $\Omega_1 = \Omega_2 = 3.186269$, Overcontact = 0.111 %**

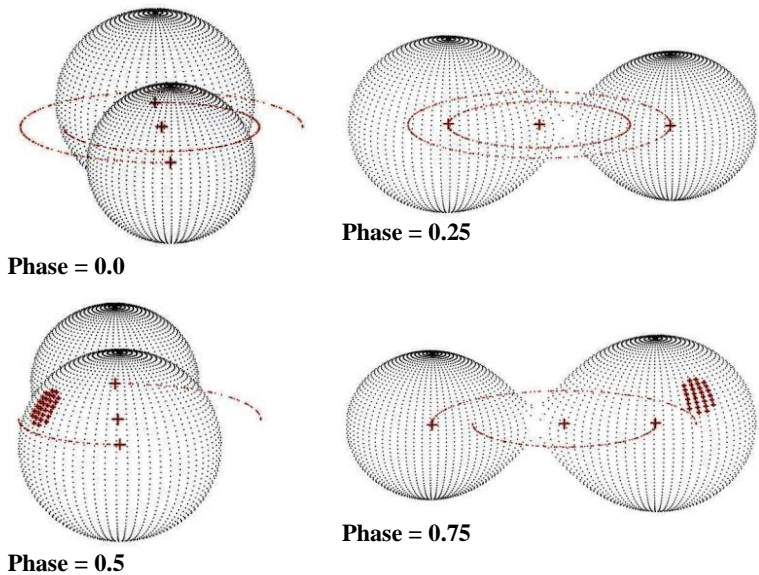

**Fig. 12: The shape of the system GSC 107-596 at phases 0.0, 0.25, 0.50, and 0.75**

## CONCLUSIONS AND FUTURE REMARKS

First photometric analyses were carried out for the systems GSC 2362-2866 and GSC 107-596. Both systems GSC 2362-2866 and GSC 107-596 exhibits slightly light variation from one side to the other (O'Connell-effect) (O'Connell 1951) in V band for the system GSC 2362-2866 and V & R bands for the system GSC 107-596 as seen from figure 1, 2, and 3 respectively. This effect can be well explained by starspot hypothesis on the components of W UMa systems that possess high surface activity because of their rapid rotation and convective envelopes.

The solutions reveal that GSC 2362-2866 is an over contact binary system by 12%, while the GSC 107-296 is an over contact binary system by 16% in V-band and 11% in R-band, which mention to a shallow common convective envelope in both systems that may be responsible for its high activity.

For the contact binary systems GSC 2362-2866 and GSC 107-596, the categorization of the two systems may like to be an A- subtype of W UMa system.

A more multi wavelength and radial velocity observations for both systems are needed to determine the absolute parameters, to allow a better description for evolutionary states of each system.